# Implementation of Digital Circuits on Three-Dimensional FPGAs Using Simulated Annealing

Hemin Rahimi
department of electrical engineering
university of kurdistan
Sanandaj, Iran
hemn.rahimi@uok.ac.ir

Hadi jahanirad
department of electrical engineering
university of kurdistan
Sanandaj, Iran
h.jahanirad@uok.ac.ir

*Abstract*—3D FPGAs have recently been produced as the next generation of the FPGA family to continue the integration of more transistors on a single chip seamlessly. In this paper, we propose a complete CAD flow to implement an arbitrary logic circuit on the 3D FPGA. The partitioning and placement stages of the flow are based on the simulated annealing algorithm. Furthermore, the routing stage is a modified version of the *Pathfinder* algorithm. The simulation results indicate that the comparison between 2D FPGA and 3D FPGA (including 2-tier) shows that the circuit speed increases by 28.66% and minimum channel width decrease by 29.92%, while the total area raises by 8.86%. Finally, the results of the comparison between 2-tier and 4-tier in 3D FPGA show that circuit speed and minimum channel width increase by 15.95% and 15.92% in 4-tier, respectively. Meanwhile, the total area increases only by 1.96%.

I. INTRODUCTION

The conventional 2D integration circuits implementation reaches its maximum capabilities, as well as the current technologies, are facing new challenges in terms of speed, area, power consumption, and reliability [3-5]. Moreover, Moore's law [2] is no longer applicable due to the physical limitations in scaling down the transistor feature size. Three dimensional integrated circuits are an attractive technology to overcome these problems. The main idea in 3D integration is stacking the dies and connecting them using vertical interconnections, which are called TSVs (Through Silicon Vias). The potential benefits of 3D integration include reduction in power consumption, delay, IC's footprint, and cost compared to conventional 2D integration [6].

FPGAs (Field Programmable Gate Arrays) have been widely used in electronic boards of various engineering systems, Due to their re-configurability. But the gap between ASICs (Application Specific Integrated Circuits) and FPGAs with current technology (conventional 2D integration) is large (over ten times less efficient in logic density, over three times worse in delay, and over three times higher in power consumption) [7]. The most significant reason for this large gap is programmable routing elements in the FPGA, which makes the reconfigurability for the user to implement the required circuit on the FPGA. On the other hand, by stacking several dies in 3D FPGA, another connection type would be added to the routing network: *vertical interconnection.*

The size of the TSV pitch is much bigger than the width of the other wires and logic gates. Consequently, the number of TSVs must be restricted to control the related area overhead. On the other hand, TSVs are used to transfer heat trapped between layers very efficiently, so the IC designers must place a sufficient number of the TSVs inside the chip.

In this paper, we present an efficient computer aided design (CAD) tool for the implementation of logic circuits on 3D FPGAs. The major parts of the proposed CAD tool are partitioning, placement, and routing. In the partitioning phase, the logic circuit would be partitioned into several parts, each of them is implemented on a layer. Then each part is placed and routed on a single layer, along with the interconnections among partitions that are realized using TSVs. Our developed tool is inspired by the versatile place and route flow (VTR) which is used for the circuit's implementation on 2D FPGAs [8]. Authors in [1] proposed a three-dimensional place and route (TPR) CAD flow for FPGAs wherein hMetis [9] is applied for circuit partitioning. The TPR considers a symmetric 3D architecture for all switch boxes, which means there is an equal number of wire tracks on the x-axis, y-axis, and z-axis. The tracks in the z-axis are realized using TSVs with a very large size in comparison to x and y wire tracks. On the other hand, only a low fraction of TSVs are used to implement vertical interconnection in practice. Consequently, using 3D switch boxes would result in a large area overhead along with a high cost for the fabrication of a large number of unnecessary TSVs. This feature makes the chip not optimal in terms of cost, area, latency, and power consumption. Moreover, the TPR router does not support timing-driven.

SA (Simulated Annealing) is one of the key meta-heuristic algorithms to approximate global optimization in a large search space and nonlinear problems [10]. The SA algorithm has been efficiently applied to solve some important problems such as Travelling Salesman Problem [11], the packing problem [12], Timetabling Problem [13], and Supply Chain Management [14]. Due to the implementation of very complex circuits on digital ICs, the typical optimization problems in this field are related to a very large search space, which makes SA an efficient tool for such problems. There are many academic types of research, in the field of IC design, that exploit SA [15-19].

In our proposed CAD tool, we have applied the SA algorithm in the partitioning and placement stages to find the optimum partitions and optimally place the partitions on 3D FPGA layers, respectively. Moreover, the routing stage in our proposed flow is accomplished using a modified *Pathfinder* routing algorithm. *Pathfinder* [22] uses an iterative algorithm that converges to a possible solution in which all signals are routed. This algorithm is used in related commercial (Altera, Xilinx) and academic CAD tools [1], [8], [21], [22].

The rest of this paper is organized as follows: Section II presents the preliminaries of this paper. Section III presents our proposed flow, and the experimental results are provided in Section IV. Finally, Section V summarizes the respective conclusions of this work.

## II. PRELIMINARIES

In this section, we review the basic concepts related to the architecture of 2D FPGAs and the new concepts which are arisen in 3D FPGAs, such as 3D Switch Boxes and TSVs. Then the optimization problems related to partitioning, placement, and routing stages would be defined for 3D FPGAs.

### A. 2D FPGA: architecture and components

Fig. 1 presents a top view of a 2D FPGA architecture which is called mesh-based architecture. Its main components are Configurable Logic Blocks (CLBs), Switch Boxes (SBs), Connection Blocks (CBs), and I/O Pads which are implemented on a regular grid. In the following, we briefly describe the structure of each of these components.

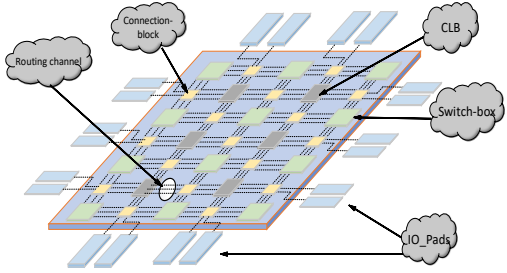

Fig. 1. A typical mesh-based 2D architecture with its main

The logical function of the implemented logic circuit would be realized using CLBs. Each CLB contains a cluster of BLEs (Basic Logic Elements) for which the local routing resources connect their input/output pins to the related CLB's ports. A BLE consists of an LUT (Look-Up Table) and a DFF. Fig 2. shows a typical BLE architecture.

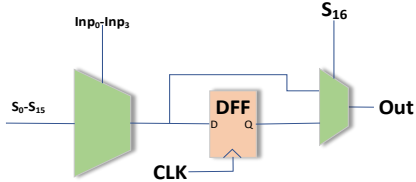

Fig. 2. A typical BLE structure (lut-4)

The routing network consists of wire tracks, Connection Blocks (CBs), Switch Boxes (SBs) and routing channels. Each CLB's input and output ports would be connected to the neighbor routing channels through adjacent CBs. On the other hand, an SB is inserted in the intersection of two horizontal and vertical channels, so that a wire track that belongs to one of these adjacent routing channels could be connected to the wires of the other channels using the SB [21].

Fig. 3a shows a 2D switch box along with its adjacent routing channels. $F_S$ and *pip junction* are two related concepts which are defined as follows; The flexibility ($Fs$) of a switch box is the total number of output tracks that connect to an input track and *pip junction* represents the sum of the entire pins at all sides of the switch box. The 2D SB of Fig. 3a has $F_s = 3$ and *pip junction*=16.

To compare the 2D switch box with the 3D switch box, suppose that *pip junction* in each side to be $W_{min}$ (minimum channel width) along with the pass transistor is used to connect the tracks of the adjacent routing channels. the total number of switches (pass transistors) in 2D and 3D SBs would be $W_{min} \times (3/2)$ and $W_{min} \times (5/2)$, respectively [22]. Furthermore, the fabrication of $W_{min}$ TSVs in upper and lower faces of SB results in a significant area overhead.

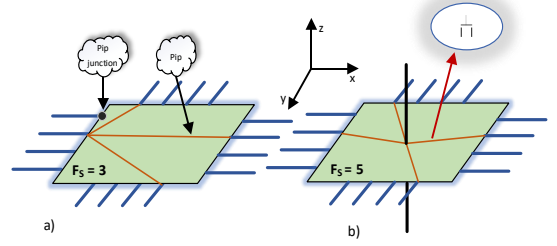

Fig. 3. FPGA Switch Box overview: a) 2D SBs, b) 3D SBs

### B. 3D FPGA Structure

Fig. 4.b shows an excellent overview of a stacked 3D FPGA architecture in four active tiers that the adjacent tiers are connected to each other. As illustrated in Fig. 4.b, within the 3D stacked model, the interconnection between the active tiers is accomplished using vertical connections called Through Silicon Vias or TSVs. As the first direct result of this action, the *critical path* length is potentially reduced. To better comprehend this, suppose a 2D chip in Fig. 4.a with Manhattan distance *2X*. Currently, divide this chip into four equal tiers and stack them tier by tier to produce a 4-tier chip (Fig. 4.b).

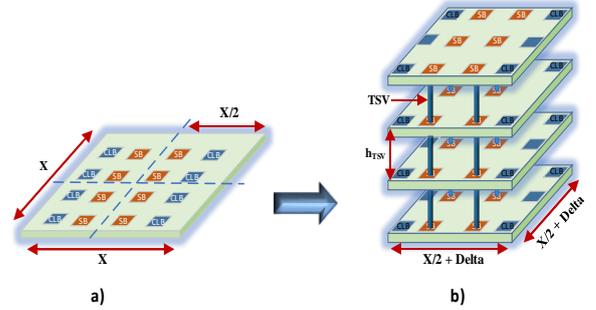

Fig. 4. a) 2D FPGA structure, b) 3D FPGA structure

Consider $h_{TSV}$ as the TSV height (It will be explained in the following subsections) and the chip dimension (length and width) was *X/2 + Delta* wherein, *Delta* is a parameter due to TSVs area overhead. Hence, the maximum Manhattan distance ($D_{max}$) is:

$$D_{max} = 2 \times (X/2 + Delta) + 3 \cdot h_{TSV} \quad (1)$$

In general, maximum Manhattan distance during a 3D chip with *n* tiers and chip dimension X and Y (length and width) is:

$$D_{max} = (X + Y) + (n - 1) \cdot h_{TSV} \quad (2)$$

Where it is assumed that area overhead due to TSVs (*Delta*) is embedded in X and Y. From Eq. (2), as expected, we will conclude that in 3D integration the critical path and wirelength decreases, and its direct effect is to reduce chip delay. Also, the area occupied by the chip is reduced and logic density increased.

### C. Through Silicon Via (TSV)

TSVs have lower power loss and circuit delay as well as higher bandwidth, in comparison with standard wire bonds [23]. Recent researches are focused on designing reliable and cost effective TSVs. The major factors which affect the design of TSV architecture are filler material (it is a material used to filling TSV), height, diameter, and shape. Cooper

(Cu), tungsten, and polycrystalline silicon are the most common filler materials. In modern designs, the TSV shape can be circular, annular, square, tapered, or rectangular and yet this is one of the critical challenges that currently exist as to which shape and for what use is most useful. Finally, to obtain an overview of the important dimensions of TSVs, consider TSV roadmaps on ITRS 2015. According to this roadmap, TSV diameter is 2-4µm, TSV pitch is 4-8µm and TSV height is 30-50µm in the time frame 2015-2018. Ultimately, the number of TSVs in 3D FPGA chips must be controlled efficiently to restrict the related area overhead as low as possible.

*D. Partitioning, Place and Route problem*

In the preliminary step of a circuit implementation, the circuit graph is generated wherein the gates and interconnection are transformed to vertices and edges, respectively. Assuming that the circuit graph is represented by $G = (V, E)$, wherein $V = \{v_1, v_2, …, v_n\}$ is the set of vertices and $E = \{e_1, e_2, …, e_n\}$ is the set of edges. In partitioning problem, we want to divide V into $k$ different subsets $V_1, V_2, V_3, …, V_k$; Where:

$|V_1| = |V_2| = … = |V_k|, V_i \cap V_j = \Phi, i \neq j$

$\bigcup_{i=1}^{k} V_i = V$

There are several solutions for graph partitioning but to reach the optimum solution a cut-size based cost function is used generally. For $k$ resulted partitions, the cut-size would be the number of edges that interconnect two vertices of two different parts. In our 3D FPGA related problem, each partition consists of the gates which would be implemented on a tier. So, the cut-size roughly represents the number of necessary TSVs. Consequently, reducing the cut-size based cost function leads to reduction of required TSVs. In this paper, we propose an SA based partitioning algorithm that is both efficient and scalable. Note that the acceptable partitioning runtime must be achieved for nowadays very large integrated circuits.

The subsequent stages after the partitioning step, are placement and routing. The optimum locations for the circuit's gates are determined in the placement phase. The placement approaches which widely used in the literature could be categorized as timing-driven placement (that is focused on maximizing circuit speed), wirelength-driven placement (Minimize overall circuit wiring), and routability-driven placement (balancing the wiring density and congestion controlling). As mentioned in the first section, the most popular and widely used algorithm in FPGAs is pathfinder [21]. The routing stage depends heavily on how the placement performs in the previous step. In other words, the overall quality of the layout, in terms of area and performance is principally determined within the placement phase [24]. In this paper, we develop the SA-based approaches for placement as well as a modified routing algorithm for 3D FPGA.

## III. PROPOSED FLOW

The overall flow of our proposed method has been illustrated in Fig 5. The 3D FPGA architecture is presented as an .xml file to the flow. Moreover, the circuit that is described in Verilog HDL would be the other input to the flow.

In the first step, the Verilog description of the circuit is synthesized, optimized, and technology-mapped based on the architecture file. We used odin_II [25] and ABC [26] for the accomplishment of this essential step. The result of this step is a gate-level description of the circuit in BLIF (Berkeley Logic Interchange Format) file format. In the 2nd step, the BLIF file is partitioned into some parts equal to the 3D FPGA's tiers. In the last step, appropriate placement and routing processes would be performed based on the proposed three-dimensional architecture.

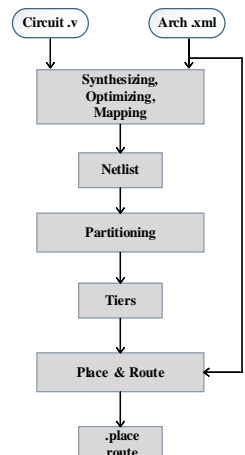

Fig. 5. Our proposed flow for 3D FPGA design

*A. Proposed 3D architecture*

our proposed architecture contains a mixture of 2D and the 3D switch boxes. Number of the 3D Switch boxes is about 33% of all programmable switches and this pattern is repeated in all tiers. The CBs and CLBs (containing an LUT with size K = 6) are fully 2D modules as well as wire segment lengths are assumed Single, Double, and Quad. The vertical channel width and the horizontal channel width ($W_{min}$) are properly selected by the router (minimum channel width required to route the circuit). The TSVs are in square shape and the related model parameters, according to [6] and [27], are shown in Table. I.

Table I. TSV parameters

| | |
|---|---|
| Resistance | 0.35 Ω |
| Capacitance | 3 fF |
| Diameter | 2 µm |
| Pitch | 4 µm |
| Height | 20 µm |

*B. Partitioning algorithm*

The first step toward circuit implementation on 3D FPGA is the partitioning in which for all tiers it would be determined which circuit gates must be placed and routed. As mentioned before, the interconnections among partitions are realized by TSVs which are restricted in our proposed 3D FPGA architecture to reduce the related area overhead. So, we have selected the cut-size as the cost function of our proposed partitioning algorithm. This cost function results in minimum TSV utilization. Furthermore, the size of the partitions in our proposed algorithm is balanced that means all partitions contain approximately equal number of gates. Our proposed algorithm (Algorithm. I) is based on simulated annealing wherein both runtime and minimum cut-size problems are handled properly.

Algorithm. I gets the circuit graph (G (V, E)) as its input, and outputs an optimized partitioning which includes $n$ sub-circuit graphs. In the first step, the circuit graph randomly is partitioned into $n$ equal-sized subgraphs (*Line 1*). Then based on *Setup()* function the initial gain would be constructed. The gain of a graph vertex is computed according to (3), wherein $E_i$ and $I_i$ are the external and external costs, respectively.

$$D_i = E_i - I_i \qquad (3)$$

The edges which are connected to a vertex ($v$) in the tier $j$'s graph are grouped into external and internal edges. The connection between $v$ and any vertex in the same tier is realized by the internal edge. On the other hand, an external edge would connect this vertex ($v$) to a vertex in the other tiers. The number of external and internal edges of a vertex would be the External and Internal costs, respectively.

In the main loop of the algorithm (Lines 6-28), for each temperature ($T$), an $N$ number of local moves would be applied. In every local move, two vertices from different tiers are selected then these vertices are swapped and the new cost function would be calculated for this configuration. If the new configuration reduces the cost, then the local move would be accepted otherwise the basic SA algorithm's approach would be applied for acceptance or rejection of the local move (Line 16-20). The related local parameters ($\alpha$ and $N$) are set using *Setting local parameters()* function as well as the pair of vertices selection for every local move would be done using *V_Selection(S')* function.

The main loop is repeated until a *stall* occurs in the cost function or $T$ reaches its minimum value ($T_{min}$). The *stall* happens when the cost function value does not change sufficiently or remains unchanged for an outsized number of local moves.

In each iteration, the temperature is updated according to the cooling schedule. The cooling schedule is presented in Eq. 4.

$$T_{(k+1)} = T_{(k)} \cdot \alpha \qquad (4)$$

Where, $k$ is iteration number and $\alpha$ is the cooling coefficient ($0 < \alpha < 1$).

In this algorithm, the cost function is presented in (5).

$$\text{Cost} = (I_i + I_j) - (E_i + E_j) + (2 \cdot C_{ij}) \qquad (5)$$

Where $I_i$ and $I_j$ are the internal costs, $E_i$ and $E_j$ are the external cost of $v_i$ and $v_j$ vertices respectively and $C_{ij}$ is the number of direct interconnections between those vertices.

when a geometric cooling schedule is employed, the high initial temperature does not considerably improve the optimization process. On the other hand, if a low value is assigned thereto, the algorithm might not achieve the optimal solution. In our algorithm, the initial temperature ($T_0$) is the cut-size of the initial solution.

Moreover, $T_{min}$ is set to 0.5 and $\alpha$ is:

$$\alpha = \frac{Cost\_final}{Cost\_initial}$$

where, *Cost_initial* is the first accepted local cost in each temperature ($T$), when $\Delta Cost < 0$. Also, *Cost_final* is the last accepted local cost in each temperature when $\Delta Cos < 0$.

**Algorithm I.** Simulated Annealing-based partitioning

**Input:** G (V, E)
**Ensure**: Optimized partitioned Graph to $l$ tiers

1: ***Randomly_partitioning( )***
2: ***Setup( )***
3: Setting parameters (T, $T_{min}$)
4: S ← $S_{init}$
5: $Cost_{Current}$ ← Number of TSVs
6: **while** T > $T_{min}$
7:    ***Setting local parameters (alpha, N)***
8:    **for** i = 0: N
9:      ***V_Selection(S')***
10:     Calculate Cost function
11:     Calculate $\Delta$Cost
12:     **if** ($\Delta$Cost < 0)
13:       S ← S'   // updating
14:       $Cost_{Current}$ ← $Cost_{current}$ + $\Delta$Cost
15:     **else**
16:       r ← Rand (0,1)
17:       **if** (r < exp(-$\Delta$Cost/T))
18:         S ← S'   // updating
19:         $Cost_{Current}$ ← $Cost_{current}$ + $\Delta$Cost
20:       **end if**
21:     **end if**
22:    **end for**
23:    **if** (*stall*)
24:      **break;**
25:    **else**
26:      T = G (T, K)
27:    **endif**
28: **end while**

*C. Placement*

The placement problem is defined as finding an optimal assignment of each vertex $V_i$ to a location such that there would be no more than one vertex in every location [24]. The more precise and better placement leads to a circuit routing with better quality.

The placement is an *NP-hard* problem, consequently, in the CAD tools, metaheuristic methods are utilized to accomplish this step [28]. In this paper, we develop an SA-based approach for the 3D FPGA placement phase. In this algorithm, first of all, a random placement is generated (for each tier), then for each temperature, a fixed number of local moves (swap the locations of two randomly selected gates) are performed. The acceptance or rejection of a local move would be done similar to the basic SA algorithm.

*D. Routing*

In the last step, the required gates would be done using routing resources. Similar to VTR and TPR CAD tools, we employ the *Pathfinder* algorithm to accomplish the routing process. The main issue in 3D FPGA routing which should be handled properly is how to includes the vertical interconnections in the *Pathfinder* algorithm. In contrast to TPR which only minimizes the wirelength in the routing step, our proposed algorithm is both timing-driven and wirelength-driven.

In the proposed routing algorithm, for each net (a source and the related sinks), a directed graph is constructed. The set of vertices $V$ in $G$ $(V, E)$, represents the I/O terminals of internal blocks (logic elements, I/O Pads, etc.) and the routing wires in the routing regions. Also, an edge between a pair of vertices represents a route between the related modules.

In our proposed routing algorithm, at the first iteration, all the nets (one by one) are routed using Dijkstra's shortest path algorithm. After this iteration, some routing resources are shared among multiple nets. In the next iterations, the related nets would be ripped-up and re-routed to solve the resource sharing problem.

## IV. RESULTS

In this section, we have used MCNC benchmarks circuits for our experiments. Furthermore, the developed algorithms have been implemented using C++ language on a Linux (Ubuntu) platform with four CPU cores (2.9 GHz) and 8GB RAM. The parameters of 2D FPGA and 3D FPGA (such as timing parameters and layouts) have been extracted from SPICE simulations based on the 22nm PTM technology [29].

Authors in [30] have developed a partitioning-based placement (Parti-SA) for 3D FPGAs in 2-tier. We compared our proposed partitioning algorithm with Parti-SA in terms of the total TSVs. As represented in Fig. 6, our proposed algorithm is more efficient and its quality (in terms of TSV reduction) improved by 9.89%.

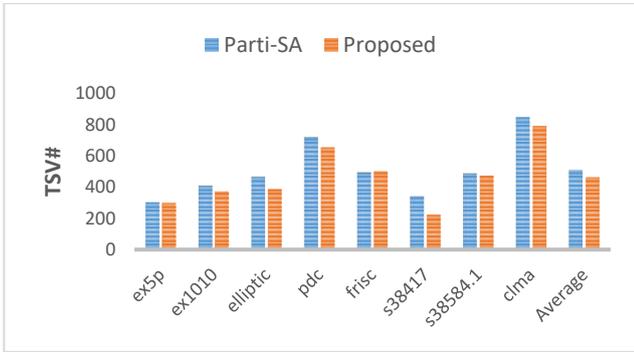

Fig. 6. Comparison between Parti-SA and the proposed algorithm in terms of TSV number.

In Table. II, we compare our algorithm with TPR CAD tool in terms of TSV Count, Circuit speed and total wirelength.

As can be seen (in the last row), the TSV number is improved by 5.34%, the circuit speed is increased by 31.14% and total wirelength is improved by 5.16%.

Table. III represents the simulation results for the comparison between 2D Design (VPR 8.0) with the results of our CAD flow for 3D Design. As can be seen in the "*IMPROVEMENT*" row, the critical path delay (CPD) improved by 28.66% in 3D Design. The total number of tracks ($W_{min}$) is reduced in 3D FPGA by 29.92% in average. Also, the average number of transistors that are added to a three-dimensional design compared to a two-dimensional design is just 8.86%. The Area term in Table. III is the total area of all tiers. We compared our method (in 2 tiers) to virtex-7 (*xc7vx1140tflg1930-1*) in synthesis stage (vivado xilinx) in terms of circuit speed. In our method the circuit speed is improved by 26.72%.

In Fig. 7, we compare critical path delay in 2-tier and 4-tier 3D FPGA. In addition, we do this comparison for the minimum channel width (Fig. 8) and transistor count (Fig. 9) parameters. As can be seen in related charts along with Table. 4, the critical path delay in the 4-tier chip decreases by 15.95% compared to the 2-tier. The enhancement is 15.92% for channel width. In this transition, the area just increases by 1.96% in 4-tier, which is negligible in comparison to the gap between two-dimensional and 2-tier three-dimensional FPGAs.

Table. II. Comparison between TPR and proposed CAD flow in terms of TSV count, critical path delay and total wirelength (WL)

| Circuit | TPR | | | Proposed | | |
|---|---|---|---|---|---|---|
| | TSV# | Delay | WL | TSV# | Delay | WL |
| ex5p | 116 | 4.26 | 12361 | 109 | 2.58 | 11657 |
| misex3 | 175 | 3.82 | 12894 | 153 | 4.31 | 11561 |
| apex2 | 192 | 5.71 | 18653 | 168 | 3.92 | 17319 |
| apex4 | 182 | 5.43 | 17069 | 146 | 5.01 | 18235 |
| alu4 | 124 | 5.67 | 11364 | 109 | 2.68 | 9987 |
| seq | 203 | 4.92 | 17246 | 172 | 3.62 | 16023 |
| des | 126 | 3.09 | 19143 | 112 | 1.93 | 17462 |
| pdc | 526 | 6.88 | 73649 | 542 | 4.07 | 71482 |
| ex1010 | 340 | 6.34 | 56173 | 314 | 4.02 | 54320 |
| elliptic | 391 | 7.58 | 38195 | 409 | 3.84 | 36412 |
| clma | 637 | 7.96 | 103742 | 617 | 5.79 | 97378 |
| **Average** | **273.82** | **5.51** | **34589** | **259.18** | **3.79** | **32894** |
| **Imp(%)** | | | | **+5.34** | **+31.14** | **+5.16** |

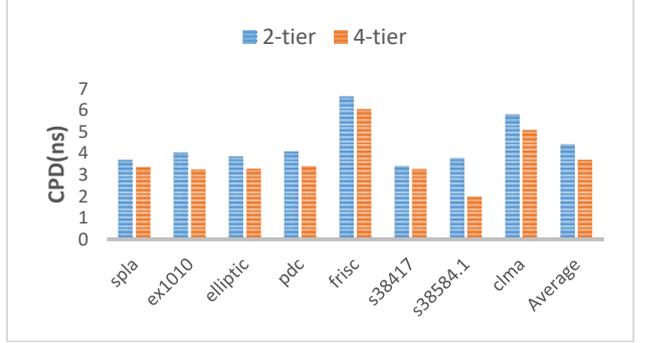

Fig. 7. comparing circuit delay between 2-tier and 4-tier

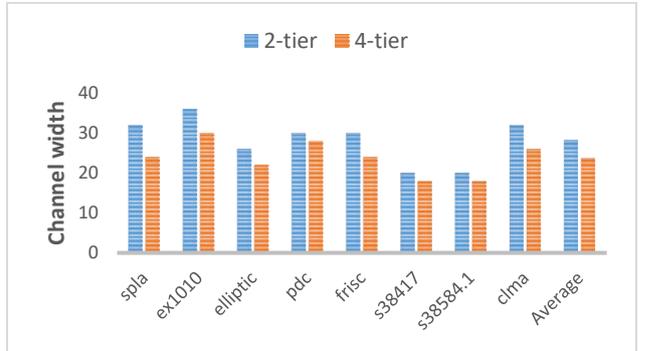

Fig. 8. comparing track number in routing channels between 2-tier and 4-tier

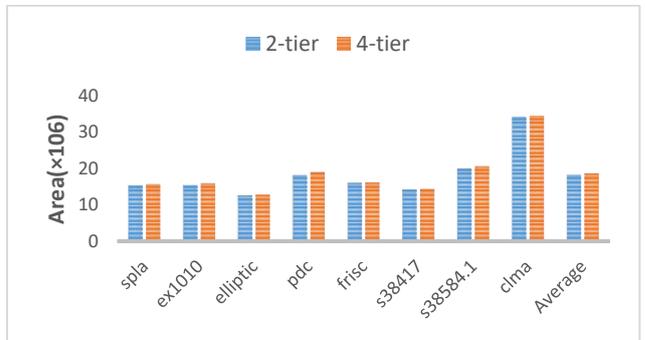

Fig. 9. Comparing transistor count between 2-tier and 4-tier

## V. Conclusion

In this paper, a complete flow has been developed for implementation of logic circuits on the 3D FPGAs. The results show that in comparison to 2D FPGA with 2-tier 3D FPGA, the important properties such as circuit delay and minimum channel width are improved by 28.66% and 29.92%, respectively. Meanwhile, the total area (all tiers) only increases by 8.86%. Also, in the comparison between 3D FPGA with 2-tier and 4-tier, the result shows that the circuit speed increases by 15.95% and the channel width decrease by 15.92% while the increase in total area is negligible. Our results show three-dimensional FPGA with four tiers is better than two tiers.

| Table. III. Comparison between 2D and 3D FPGA | | | | | | | | |
|---|---|---|---|---|---|---|---|---|
| Circuit | | ex5p | apex2 | alu4 | seq | ex1010 | elliptic | clma |
| 2D | **CPD**(ns) | 3.44 | 5.18 | 3.41 | 4.78 | 6.27 | 6.08 | 7.92 |
| | **W$_{min}$** | 24 | 34 | 28 | 36 | 48 | 36 | 48 |
| | **Area** ($\times 10^6$) | 3.16 | 6.38 | 4.79 | 5.67 | 13.96 | 11.43 | 30.94 |
| 3D | **CPD**(ns) | 2.58 | 3.92 | 2.68 | 3.62 | 4.02 | 3.84 | 5.79 |
| | **W$_{min}$** | 18 | 26 | 18 | 22 | 36 | 26 | 32 |
| | **Area** ($\times 10^6$) | 3.41 | 7.01 | 4.96 | 6.17 | 15.39 | 12.64 | 34.17 |
| **IMPROVEMENT**: **DELAY** = +28.66%     **W$_{MIN}$** = +29.92%     **AREA** = -8.86% | | | | | | | | |